\documentclass[aps,prd,letterpaper,twocolumn,showpacs,preprintnumbers]{revtex4}
\usepackage{amsmath}
\usepackage{amssymb}
\usepackage{amsfonts}
\usepackage{latexsym}
\usepackage{graphicx}
\usepackage{dcolumn}
\usepackage{bm}
\usepackage{empheq}
\usepackage{color}
\usepackage{times}  
\newcommand{\eq}[1]{\begin{equation}#1\end{equation}}

\newcommand{\lrb}[1]{\left( #1 \right)}  

\definecolor{gxhighlight}{rgb}{1,1,0.4}

\begin{document}

\preprint{CAS-KITPC/ITP-196}

\title{Testing gravity with non-Gaussianity}

\author{Xian Gao}%
    \email{xgao@apc.univ-paris7.fr}
    \affiliation{%
        Institute of Theoretical Physics, Chinese Academy of Sciences, Beijing 100190, China\\
        APC, CNRS-Universit\'{e} Paris 7, 10 rue Alice Domon et L\'{e}onie Duquet, 75205 Paris Cedex 13, France\\
        Laboratoire de Physique Th\'{e}orique, \'{E}cole Normale Sup\'{e}rieure, 24 rue Lhomond, 75231 Paris, Cedex 05, France, and\\
        IAP (Institut d'Astrophysique de Paris), 98bis Boulevard Arago, 75014 Paris, France
        }%

\date{\today}


\begin{abstract}

We show that modified gravity presents distinctive nonlinear
features on the Cosmic Microwave Background (CMB) anisotropies comparing with General Relativity (GR).
We calculate the contribution to the CMB non-Gaussianity from nonlinear Sachs-Wolfe effect in $f(R)$ gravity and show that, contrary to GR's contribution which is typically $\lesssim \mathcal{O}(1)$, the contribution in $f(R)$ gravity is sensitive to the nonlinear structure of $f(R)$ and can be large in principle.
Optimistically, this gives an alternative origin for the possibly observed large CMB non-Gaussianities besides the primordial ones.
On the other hand, such nonlinear features
can be employed to provide a new cosmological test of $f(R)$ or other modified theories of gravitation, which is unique and  independent of previously known tests.

\end{abstract}

\maketitle%


Understanding the origin and evolution of observed large-scale structure of the Universe plays a key role in cosmology. With the increasingly precise data of current and upcoming observations on cosmic perturbations (CMB, LSS), we are able to gain unprecedented understandings of both the Universe and the fundamental physics relevant to it.
An unique observable of cosmic perturbations is non-Gaussianity, which characterizes the nonlinearities of cosmic perturbations \cite{Bartolo:2004if,Lyth:2005fi}.
The possibly observed non-Gaussianities
in CMB anisotropy have two main origins: the ``primordial"
non-Gaussianities of curvature perturbation $\zeta$ \cite{Maldacena:2002vr,Koyama:2010xj}, and the post-inflationary nonlinearities in the
transfer from $\zeta$ to anisotropies, or other secondary contributions \cite{Bartolo:2010qu}.
The
former is due to the interactions in the models or mechanism which
generate the primordial perturbations, and is thus expected to encode early Universe Physics. The later,
however, which is due to the intrinsic nonlinearities in gravitation
and photon-baryon plasma, is often referred to as ``contaminations"
to the primordial non-Gaussianity, since the physics of CMB is
well-understood and they are too small to
be detected, especially compared with their possibly ``large"
primordial partners \cite{Bartolo:2010qu,Bartolo:2004ty}.

In this Letter, we investigate the ability of these post-inflationary contributions to non-Gaussianities to probe gravitational theories alternative to General Relativity (GR). In the past several years, modified gravity theories (see e.g. \cite{Sotiriou:2008rp} for reviews) have gained much attention due to the unexpected later-time cosmic acceleration, which are employed as an alternative to dark energy models \cite{Carroll:2003wy,Hu:2007nk}. Most of  existed cosmological constraints to modified gravity theories (see e.g. \cite{Jain:2010ka} for a recent review) come from the observation of CMB \cite{Song:2007da,Song:2006ej}, LSS \cite{Tsujikawa:2007tg}, weak lensing \cite{Schmidt:2008hc}, solar system test \cite{Will:2001mx} as well as correction to the Newton's law \cite{Nojiri:2007jr}.
In this Letter, we propose a new observational window to test modified gravity, i.e. the nonlinear anisotropy due to gravitational perturbations \cite{Pyne:1993np,Bartolo:2010qu,Bartolo:2004ty}.
As is well known, the large-scale radiation transfer comes from two aspects: the intrinsic anisotropy on the emission surface and the gravitational redshift undergone by the photon from the emission surface to us. The later is kinetic and is irrelevant to the gravitational theory. However, the former (as well as the nonlinear relation between metric perturbations and conserved curvature perturbation $\zeta$) depends on the underlying gravitational theory. Thus, intuitively, one may use this nonlinear transfer to probe the gravitational theories.
In this Letter we focus on the simplest metric $f(R)$ gravity.
For the first time, we give the nonlinear generalization of the (ordinary) Sachs-Wolfe
effect in $f(R)$ gravity, up to the second order.
Analogous to what happens in $k$-inflation model with matter Lagrangian $P(X,\phi)$ where
primordial non-Gaussianities are enhanced by the nonlinear dependence of $P$ on $X$ \cite{Seery:2005wm},
 in $f(R)$ gravity the nonlinear transfer from primordial curvature perturbation into
anisotropy is controlled by the nonlinear structure of $f(R)$, and can also be enhanced in principle.


We investigate a class of metric $f(R)$ gravity with  action
    \eq{
    S= \int d^4x \sqrt{-g}\lrb{\frac{1}{2} f(R) + \mathcal{L}_m},
    }
where we set $8\pi G=1$.
The equation of motion for $g_{\mu\nu}$ is
    \begin{equation}{\label{GEE}}
        \Sigma_{\mu\nu}\equiv
        f_{,R}R_{\mu\nu}-\frac{1}{2}f\left(R\right)g_{\mu\nu}+\left(g_{\mu\nu}\square-\nabla_{\mu}\nabla_{\nu}\right)f_{,R} = T_{\mu\nu},
    \end{equation}
where $f_{,R}\equiv df/dR$ etc. and $\square \equiv
\nabla_{\mu}\nabla^{\mu}$.

After decoupling, the CMB photon density remains as Planck
distribution, which implies ${T_f}/{T_i} = {\omega_f}/{\omega_i}$,
where $T_f$ ($T_i$) and $\omega_f$ ($\omega_i$) are the final (initial) temperature and photon energy respectively.
The temperature anisotropy is thus given by
    \eq{{\label{anisotropy_general}}
        \frac{\Delta T}{T} \equiv
        \frac{T_o-\bar{T}_o}{\bar{T}_o}
        =\frac{a_{o} \omega_{o}}{a_{e} \omega_{e}} \frac{T_e}{\bar{T}_e}-1
        ,
    }
where
$\omega_o$ ($\omega_e$) are scale factor and photon energy at the observer (emission surface).
From (\ref{anisotropy_general}) it is clear that the observed
anisotropy has two contributions: one is the gravitational
redshift-induced anisotropy  ${a_o\omega_o}/{(a_e\omega_e)}$, which is purely kinetic, the other is the intrinsic anisotropy $T_e/\bar{T}_e$
on the emission surface which depends on the dynamics during
recombination and the gravitational theory.


For later convenience, we introduce
    \begin{equation}{\label{epsilon_def}}
      \epsilon = 1-\frac{d\ln\mathcal{H}}{d\ln a},\quad \epsilon_{R}=\frac{d\ln R}{d\ln a},\quad\epsilon_{R'}=\frac{d\ln{R}'}{d\ln a},
    \end{equation}
where $R'$ denotes $\partial R/\partial \eta$ with $\eta$ the comoving time.
It is also useful to employ a set of dimensionless ratios, which characterize the nonlinear structure of $f(R)$:
    \begin{equation}{\label{ratio_def}}
      \beta=\frac{Rf_{,RR}}{f_{,R}},\quad\gamma=\frac{R^{2}f_{,RRR}}{f_{,R}},\quad \delta=\frac{R^{3}f_{,RRRR}}{f_{,R}}.
    \end{equation}
Recall that in Einstein gravity, we have $\epsilon =3(1+w)/2$ where $w$ is the constant equation of state parameter, and $\epsilon_R=-2\epsilon$, $\epsilon_{R'}=1-3 \epsilon $, $\beta=\gamma=\delta=0$. The ``Compton parameter" $B$ introduced in \cite{Song:2006ej} which was employed by several authors can be expressed as $B=-\beta\epsilon_{R}/\epsilon$, which should satisfies $B \geq 0$ for the stability of linear perturbations \cite{Song:2006ej}.

We focus on the ordinary Sachs-Wolfe contributions, which dominate on the large scales. Thus we
 consider the scalar metric perturbations in Newtonian gauge
    \begin{equation}{\label{metric_pert}}
      ds^2 = a^{2}\left(-e^{2\Phi}d\eta^{2}+e^{-2\Psi}dx^{i}dx^{i}\right).
    \end{equation}
As mentioned before, the gravitational redshift contribution to the observed anisotropy is the same for all metric theory of gravitation, which is, up to the second order in $\Phi$ \cite{Pyne:1993np,Bartolo:2010qu,Bartolo:2004ty,Bartolo:2005fp}
    \begin{equation}{\label{redshift}}
      a\omega = -\Phi + \frac{1}{2}\Phi^2.
    \end{equation}
What is interesting is the intrinsic temperature anisotropy  on the emission surface. Assuming the cosmic perturbation is adiabatic, in matter dominated era, the large-scale temperature fluctuation has the relation: $ T_{e}/\bar{T}_{e}=\left(\rho_{\gamma}/\bar{\rho}_{\gamma}\right)^{\frac{1}{4}}=\left(\rho_{m}/\bar{\rho}_{m}\right)^{\frac{1}{3}}$. To determine the matter density perturbation, we have to make use of the generalized Einstein equation (\ref{GEE}).
In GR, on large scales it yields a non-perturbative relation \cite{Bartolo:2005fp}: $\rho_m/\bar{\rho}_m=e^{-2\Phi}$. In $f(R)$ gravity, up to the second-order in metric perturbations, we find
    \begin{equation}{\label{rho}}
      \ln\frac{\rho_{m}}{\bar{\rho}_{m}}=-\left(2-\sigma_{1}\right)\Phi+\frac{1}{2}\left(\sigma_{2}-\sigma_{1}^{2}\right)\Phi^{2},
    \end{equation}
where $\sigma_1$ and $\sigma_2$ are combinations of derivatives of $f$ with respect to $R$,
    \begin{eqnarray}
    \sigma_{1} & = & 2-6\frac{1+\beta(\epsilon-1)+(2\beta+\gamma)\epsilon_{R}}{2\epsilon-\gamma\epsilon_{R}^{2}-\beta\epsilon_{R}\left(\epsilon_{R'}-2\right)},\label{sigma1}\\
    \sigma_{2} & = & 4+12\frac{\beta-1+\gamma(\epsilon-1)+(3\gamma+\delta)\epsilon_{R}}{2\epsilon-\gamma\epsilon_{R}^{2}
    -\beta\epsilon_{R}\left(\epsilon_{R'}-2\right)},\label{sigma2}
    \end{eqnarray}
where parameters are defined in (\ref{epsilon_def})-(\ref{ratio_def}).
It is useful to note that in GR, $\sigma_1 = \sigma_2 = 0$, thus the familiar relation $\ln(\rho_{m}/\bar{\rho}_{m})=-2\Phi$ is recovered (up to the second order).
The intrinsic temperature fluctuation on  emission surface now takes the form:
    \begin{equation}{\label{intrinsic}}
      \frac{T_e}{\bar{T}_e}  = 1+\frac{1}{3}\left(-2+\sigma_{1}\right)\Phi+\frac{1}{18}\left(4-4\sigma_{1}-2\sigma_{1}^{2}+3\sigma_{2}\right)\Phi^{2}.
    \end{equation}
Finally, collecting (\ref{redshift}) and (\ref{intrinsic}), after some manipulations, we get the nonlinear Sachs-Wolfe contribution to the large anisotropy up to the second-order in $\Phi$ in $f(R)$ gravity:
    \begin{equation}{\label{NL_SW}}
      \frac{\Delta T}{T} = \frac{1}{3}\left(1+\sigma_{1}\right)\Phi+\frac{1}{18}\left(1+2\sigma_{1}-2\sigma_{1}^{2}+3\sigma_{2}\right)\Phi^{2}.
    \end{equation}
In the above, quantities are evaluated at the emission surface. (\ref{NL_SW}) generalizes the familiar results $\frac{\Delta T}{T}=e^{\Phi/3}$ in GR \cite{Bartolo:2005fp}.

Our next goal is to relate the observed temperature anisotropy  to the primordial curvature perturbation
$\zeta$, which encodes the information in the very early universe
and is the most frequently used variable in evaluating the
primordial non-Gaussianities in the literature. It is well known
that on large scales and for adiabatic perturbation, there is a
non-perturbative and conserved quantity which can be identified as
nonlinear curvature perturbation in uniform-density slices
\cite{Salopek:1990jq} (see also \cite{Malik:2003mv}),
defined as
    \eq{{\label{zeta_def}}
        \zeta \equiv -\Psi
        +\frac{1}{3}\int_{\bar{\rho}}^{\rho}\frac{d\tilde{\rho}}{3(\tilde{\rho}+\tilde{p})}.
    }
The conservation of $\zeta$ relies on the fact that the energy-momentum tensor of the perfect fluid is conserved, and is irrelevant to the gravitational theory.
Our next task is to set the initial conditions for
$\Phi$ and $\Psi$ on the emission surface up to the second-order
in $\zeta$.
In matter-dominated era ($p=0$),  (\ref{zeta_def}) can be integrated
to give
    \begin{equation}{\label{zeta_matter}}
        \zeta = -\Psi +\frac{1}{3}\ln\frac{\rho_{m}}{\bar{\rho}_{m}} \,.
    \end{equation}
Together with (\ref{rho}), it will gives $\zeta$ in terms of metric perturbations $\Phi$ and $\Psi$: $\zeta=\zeta[\Phi,\Psi]$.

The next step is to determine the constraint between $\Phi$ and $\Psi$ in $f(R)$ gravity.
As in GR, this can be done by employing the spatial traceless part of the generalized
Einstein equation. Applying $\partial_i\partial_j$ on both sides of $\Sigma_{ij}^{\textrm{TF}}=T^{\textrm{TF}}_{ij}$, where $\Sigma_{ij}$ is defined in (\ref{GEE}),
we are able to solve $\Psi$ in terms of $\Phi$ on large scales, which gives, at linear order
    \begin{equation}{\label{psi1_sol}}
      \Psi_{1}=\left(1-2\beta\right)\Phi,
    \end{equation}
where $\beta$ is defined in (\ref{ratio_def}).
(\ref{psi1_sol}) is the well-known result, which implies that in $f(R)$ gravity, $\Phi\neq\Psi$ even at the linear order for perfect fluid.
At the second-order, we find
    \begin{eqnarray}
        \Psi_{2}=K_{2}\left[\Phi\right] & \equiv & \partial^{-4}[\left(3\lambda-2\beta+8\beta^{2}+4\gamma\right)\left(\partial^{2}\Phi\right)^{2}\nonumber \\
         &  & +\left(\lambda+6\beta-4\beta^{2}+8\gamma\right)\left(\partial_{i}\partial_{j}\Phi\right)^{2}\nonumber \\
         &  & +4\left(\lambda+2\beta+4\gamma\right)\partial_{i}\Phi\partial_{i}\partial^{2}\Phi\nonumber \\
         &  & +4\left(\beta-\beta^{2}+\gamma\right)\Phi\partial^{4}\Phi], \label{Psi2_sol}
         \end{eqnarray}
with
    \begin{equation}{\label{lambda_def}}
        \lambda =  1+\frac{2\left(1-\beta+(3\beta/2+\gamma)\epsilon_{R}\right){}^{2}}{2\epsilon+\beta\epsilon_{R}\left(2-\epsilon_{R'}\right)-\gamma\epsilon_{R}^{2}},
        \end{equation}
where various parameters are defined in (\ref{epsilon_def})-(\ref{ratio_def}).
(\ref{psi1_sol})-(\ref{Psi2_sol}) generalize the corresponding constraints in GR \cite{Bartolo:2010qu,Bartolo:2004ty}. Note the above parameters are all evaluated on the background. The GR results corresponding to (\ref{psi1_sol})-(\ref{Psi2_sol}) can be easily recovered after plugging the GR values of parameters.
Having these relations in hand, now it is straightforward to get the initial conditions for $\Phi$ up to the second order in $\zeta$:
    \begin{eqnarray}
        \Phi_{1} & = & -\frac{3\zeta}{5-6\beta-\sigma_{1}},\label{Phi_1}\\
        \Phi_{2} & = & \frac{9\left[\left(\sigma_{2}-\sigma_{1}^{2}\right)\zeta^{2}-6K_{2}\left[\zeta\right]\right]}{2\left(5-6\beta-\sigma_{1}\right)^{3}},\label{Phi_2}\end{eqnarray}
where $K_2[\cdot]$ is defined as in (\ref{Psi2_sol}).

\begin{figure}[h]
        \centering
        \includegraphics[width=8.6cm]{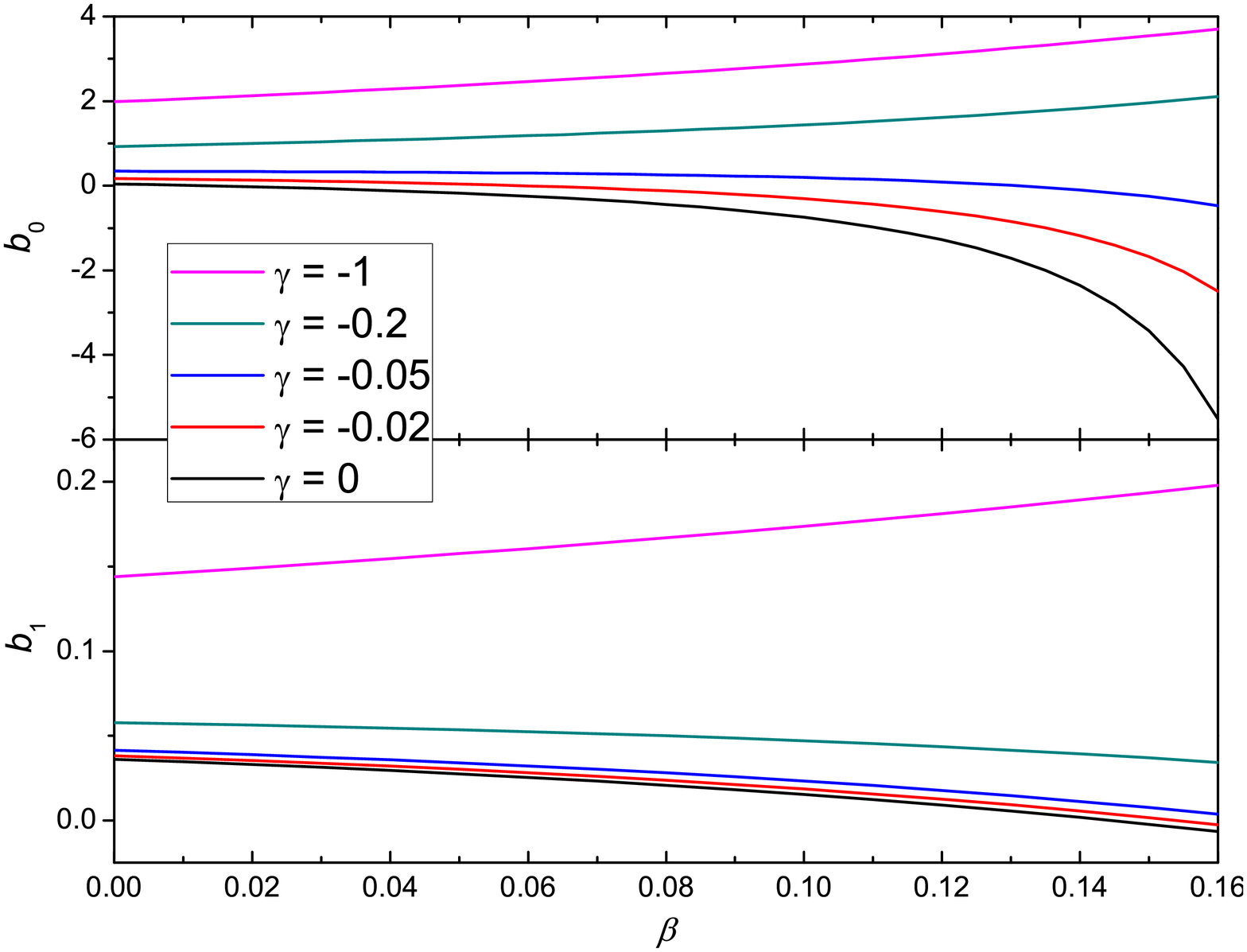}
        \caption{(Color online). $b_0$ and $b_1$ as functions of $\beta$ with diverse values of $\gamma$ and vanishing $\delta$ for the $\Lambda$CDM expansion history with $\Omega_{\Lambda}=0.74$. In the matter-dominated era, the parameters introduced in (\ref{epsilon_def}) are $\epsilon=1.5$, $\epsilon_R=-3$ and $\epsilon_{R'}=-3.5$ respectively. We assume the range of values of $\beta$, $\gamma$ and $\delta$ ensures such an expansion history.}
        \label{fig:b}
    \end{figure}

From the nonlinear Sachs-Wolfe expression (\ref{NL_SW}) and the exact expression for curvature perturbation in matter era (\ref{zeta_matter}), together with (\ref{Phi_1})-(\ref{Phi_2}), a nonlinear mapping from the conserved curvature perturbation to the temperature anisotropy can be got: $\frac{\Delta T}{T}=\left(\frac{\Delta T}{T}\right)_{\left(1\right)}+\left(\frac{\Delta T}{T}\right)_{\left(2\right)}+\cdots$, which at the linear order is
    \begin{equation}{\label{DeltaT1}}
      \left(\frac{\Delta T}{T}\right)_{\left(1\right)}=-\frac{1+\sigma_{1}}{5-6\beta-\sigma_{1}}\zeta,
    \end{equation}
and at the second-order
    \begin{equation}{\label{DeltaT2}}
        \left(\frac{\Delta T}{T}\right)_{\left(2\right)}\left(\bm{k}\right)=\frac{1}{2}\int\frac{d^{3}p_{1}d^{3}p_{2}}{\left(2\pi\right)^{3}}b\left(\bm{p}_{1},\bm{k}-\bm{p}_{1}\right)\zeta_{\bm{p}_{1}}\zeta_{\bm{k}-\bm{p}_{1}},
    \end{equation}
where
    \begin{equation}{\label{b_str}}
        b\left(\bm{p}_{1},\bm{p}_{2}\right)=b_{0}-b_{1}g\left(\bm{p}_{1},\bm{p}_{2}\right),
    \end{equation}
with momentum-independent coefficients
    \begin{eqnarray}
b_{0} & \equiv & \frac{1}{\left(5-6\beta-\sigma_{1}\right)^{3}}\Big[5-6\beta\left(7+2\left(4-\sigma_{1}\right)\sigma_{1}+3\sigma_{2}\right)\nonumber \\
 &  & \qquad\qquad\qquad+36\left(\beta^{2}-\gamma\right)\left(1+\sigma_{1}\right) \nonumber\\
 && \qquad\qquad\qquad+9\sigma_{1}-\sigma_{1}^{2}\left(15+\sigma_{1}\right)+18\sigma_{2}\Big],\label{b0}\\
b_{1} & \equiv & \frac{3\left(1+\sigma_{1}\right)\left(3\lambda+6\beta(2\beta-1)\right)}{2\left(5-6\beta-\sigma_{1}\right)^{3}},\label{b1}\end{eqnarray}
and momentum-dependent factor
    \begin{equation}{\label{g_def}}
        g\left(\bm{p}_{1},\bm{p}_{2}\right)=1+2\frac{p_{1}^{2}+p_{2}^{2}}{\left(\bm{p}_{1}+\bm{p}_{2}\right)^{2}}-3\frac{\left(p_{1}^{2}-p_{2}^{2}\right)^{2}}{\left(\bm{p}_{1}+\bm{p}_{2}\right)^{4}}.
    \end{equation}
Note in (\ref{b0})-(\ref{b1}), parameters are evaluated at the time of emission (e.g. decoupling).
The above expression has been arranged in such form so that $g\left(\bm{p}_{1},\bm{p}_{2}\right) \rightarrow 0$ as $p_1\rightarrow 0$ or $p_2\rightarrow
0$, which denotes the so-called ``squeezed" limit of momenta configuration for the bispectrum. Thus in this ``squeezed limit", $b(\bm{p}_{1},\bm{k}-\bm{p}_{1}) \rightarrow b_0$. A large $b_0$ implies a large contribution to the ``local"-type non-Gaussianity. The functional form of $b(\bm{p}_{1},\bm{p}_{2})$ is rather general, the contributions in Einstein gravity and $f(R)$ gravity differ with each other only in the different coefficients $b_0$ and $b_1$. In General Relativity, it is easy to verify that \cite{Bartolo:2010qu,Bartolo:2004ty}
    \begin{equation}{\label{b_GR}}
        b(\bm{p}_{1},\bm{p}_{2}) \xrightarrow[]{\textrm{GR}} \frac{1}{25} - \frac{3}{50} g(\bm{p}_1,\bm{p}_2),
    \end{equation}
with the same momentum factor $g(\bm{p}_1,\bm{p}_2)$. The GR expression (\ref{b_GR}) implies that, due to the smallness of the numerical coefficients, the nonlinear mapping
(\ref{DeltaT2}) would not contributes significantly, especially
in the so-called ``squeezed" momenta configurations of
non-Gaussianity
\cite{Bartolo:2005fp}, which is a long well-know conclusion in GR. In $f(R)$ gravity, however, the coefficients $b_0$ and $b_1$ depend on the nonlinear functional structure of $f(R)$, and thus can be numerically large in principle.
In Fig.\ref{fig:b}, we plot the dependence of $b_0$ and $b_1$ on parameters $\beta$, $\gamma$ for a $\Lambda$CDM expansion history  for illustrative purpose.

Most efforts of investigating the non-Gaussianity focus on the primordial ones, since they are expected to encode the early Universe physics. In this Letter, we explore the ability of ``non-primordial" (more precisely,  ``post-inflationary") non-Gaussianity to probe gravitational theories. In the context of $f(R)$ gravity, we calculate the nonlinear CMB temperature anisotropy in the Sachs-Wolfe realm. We derive the explicit nonlinear mapping from the primordial curvature perturbation to the observed anisotropy, up to the second order. We find that the amplitude of this nonlinear mapping, which is typically $\lesssim\mathcal{O}(1)$ in GR \cite{Pyne:1993np,Bartolo:2010qu,Bartolo:2004ty}, is controlled by the functional structure of $f(R)$ and thus can be large in principle, as the primordial non-Gaussianities can be enhanced by the non-canonicality of kinetic term(s) of inflaton field(s).
In this Letter we focus on the Sachs-Wolfe contributions which dominate on the large-scales.
A full treatment of the
nonlinear transfer on all scales involves solving  the full
Boltzmann equations \cite{Bartolo:2006cu}, which we leave for future investigations.

The result presented in this Letter has at least two implications. On
one hand, it gives an alternative origin to
the possibly observed non-Gaussianity --- especially the so-called ``local type" --- beyond primordial ones. On the other
hand and might be the most interesting, this sensitivity of nonlinear transfer to the structure of $f(R)$  can be employed to provide a
new test of $f(R)$ gravity, which is unique and independent of previously known tests, e.g. from spectra of CMB or LSS \cite{Song:2007da,Song:2006ej,Tsujikawa:2007tg,Schmidt:2008hc}. As has been emphasized, the momenta-dependent factor (\ref{g_def}) is universal for $f(R)$ theories but is also distinctive comparing to momentum shape functions of primordial bispectrum \cite{Seery:2005wm}. Thus, by constructing appropriate templates, we are able to disentangle $f(R)$ contributions from other primordial/non-primordial sources in the finally observed non-Gaussian signals.
 In light of current and upcoming observational constraints
on non-Gaussianities, we may have already been able put new constraints on $f(R)$ or more general modified theories of gravitation.

I thank Robert Brandenberger  for helpful discussion and
 Cyril Pitrou for illuminating correspondence. I am deeply grateful to Prof. Miao Li for
  consistent
encouragement and support. This work was partly supported by the NSFC grant
No.10535060/A050207, a NSFC group grant No.10821504 and Ministry of
Science and Technology 973 program under grant No.2007CB815401.




\end{document}